\documentclass[10pt,letterpaper,superscriptaddress,amsmath,amssymb]{article}

\usepackage{opex3}
\usepackage{bm}% bold math

\begin{document}
\title{Rigorous criterion for characterizing correlated multiphoton emissions}

\author{Hyun-Gue Hong$^1$, Hyunchul Nha$^{2,\dag}$, Jai-Hyung Lee$^1$, and Kyungwon An$^{1,*}$}
\address{$^1$Department of Physics and Astronomy, Seoul National
University, \\Seoul 151-747, Korea}
\address{$^2$Department of Physics, Texas A\&M
University at Qatar, Education City, PO Box 23874, Doha, Qatar}
\email{hyunchul.nha@qatar.tamu.edu$^\dag$}
\email{kwan@phya.snu.ac.kr$^*$}

\begin{abstract} Strong correlation of photons, particularly in the single-photon regime, has recently been exploited for various applications in quantum information processing. Existing correlation measurements, however, do not fully characterize multi-photon correlation in a relevant context and may pose limitations in practical situations. We propose a conceptually rigorous, but easy-to-implement, criterion for detecting correlated multi-photon emission out of a quantum optical system, drawn from the context of wavefunction collapse.
We illustrate the robustness of our approach against experimental limitations by considering an anharmonic optical system.\end{abstract}

 % REPLACE WITH CORRECT OCIS CODES FOR YOUR ARTICL

\section{Introduction}

% correlation between single photons and it applications

Strong correlation of photons at the few quanta level can make
possible a variety of nonlinear optical devices useful for quantum
information processing, such as single-photon transistors or
switching devices \cite{Chang2007, Faraon2008} and the generation of
single photons on demand \cite{Birnbaum2005, Dayan2008}.
Furthermore, the photon-photon correlation mediated by the emitters
can also be employed to simulate quantum many-body systems in a
controllable way. For example, the effective on-site repulsion
between photons can be exploited to study quantum phase transitions
such as the Mott-superfluid transition
\cite{Hartmann2006,Greentree2006,Angelakis2007,Na2008} and the
fermionization of bosons \cite{Chang}. These applications are
closely related to a specific correlation effect, namely a
single-photon blockade effect
\cite{Faraon2008,Birnbaum2005,Imamoglu1997,Carmichael1992}
---Once  a system is excited by one photon, the abosorption of next photons is blocked,
e.g., due to the anharmonic energy level structure of the system.

%correlation among multi-photons and its importance

Recently, interest in the correlation effect has also been extended
to multi-photon level in the context of multi-photon gateway, where
a random (Poissonian) stream of photons can be converted into a
bunch of temporally correlated $n$ photons. In particular, Kubanek
{\it et al.} demonstrated the operation of two-photon gateway to
some extent using an optical cavity QED system \cite{Kubanek2008}.
In view of all these efforts, it seems very crucial to have a
theoretical framework that can appropriately characterize
multi-photon correlations \cite{Schuster2008, Fink2008,
Bishop2008,Horvath1999}, e.g., $n$-photon blockade effect, and
desirably that can be efficiently tested in experiment.

% problem for identifying n-photon correlation

Conventionally, correlation of photons is measured by the
$n$th-order coherence functions introduced by Glauber
\cite{Glauber}, $g^{(n)}(0)=\langle a^{\dagger
n}a^n\rangle/\langle a^{\dagger}a\rangle^n$, where $a$ ($a^\dag$)
is the annihilation (creation) operator of an optical field. However, it is noted in the
recent experiments of cavity QED \cite{Faraon2008, Birnbaum2005,
Kubanek2008} that $g^{(2)}(0)$ is not effective to resolve
the correlated two photon emission due to a huge bunching at the
atom-cavity bare resonance overshadowing the two-photon resonance.
Furthermore, as $n$ goes beyond two, $g^{(n)}(0)$ or its simple variants
\cite{Kubanek2008} contain more peaks at $k=1,\cdots,n-1$ photon resonances,
irrelevant to genuine $n$-photon correlation, as to be shown below.

% previous attempts and necessity of rigorous approach

Instead one may take $n$-photon excitation peaks in $\langle a^{\dagger
n}a^n\rangle$ spectrum itself, e.g. in
\cite{Kubanek2008,Schuster2008,Fink2008,Bishop2008,Horvath1999}, as
a confirming evidence of $n$-photon correlations. Rigorously
speaking, however, the multi-photon resonant excitation peaks spectroscopically
identified only uncover the energy-level structure of the system. Whether each peak in the bare coincidence $\langle a^{\dagger
n}a^n\rangle$ indicates relevant photon correlation must be checked very carefully.
For example, the three-photon
coincidence tends to increase in the spectrum without any correlation if the system possesses two-photon
correlation since an uncorrelated emission added to a correlated
pair may register another three-photon coincidence. Therefore, we need to consider
a stricter physical context for characterizing correlated emission of $n$-photons. Although there have been several studies on
higher-order photon statistics in view of nonclassicality  \cite{Lee1990, Klyshko1996}, none of them carries a clear
interpretation as multiphoton correlation.

% our strategy and its advantages

Here we propose a conceptually rigorous, but easy-to-implement, criterion for measuring correlated multi-photon emissions out of a quantum optical system.
The criterion is derived by considering wavefunction collapse related to sequential photo-detections \cite{Carmichael1991} and has the following merits.
(i) It only detects highly-correlated $n$-photon emissions with no classical analogue, and
(ii) can be tested by usual photon coincidence detections (no conditional measurement) in a detector-efficiency insensitive form, and therefore, experimentally favorable.
We also introduce a quantitative measure of $n$-photon correlation based on this criterion, which is quite robust in addressing multi-photon correlations against experimental imperfections. This is a practical merit of our approach made possible by the well-established correlation context.
We illustrate the power of our method in an optical cavity QED system,
where genuine $n$-photon correlated emissions can be efficiently verified in accordance with its anharmonic energy levels.

\section{The criterion}

In order to envision a generic, though not
exhaustive, scenario where multiphoton correlations may arise, let
us compare two systems, one with harmonic and the other with
anharmonic level structure [Fig.1 (a)]. When a harmonic system with
level spacing $\hbar\omega_0$ is excited by an external driving on
resonance ($\omega_L=\omega_0$), all energy levels are equally
accessible. On the other hand, for an anharmonic system, if the external field is $n$-photon resonant with the $n$th level, other levels than $n$th would not be substantially addressed by the
external field. As a result, the system could be excited to contain
only $n$ correlated quanta and further excitation would be
prohibited---$n$-quanta (photon) blockade effect. We will apply a
similar line of reasoning to emission, rather than excitation,
process. Specifically, we construct a criterion to detect `pure'
$n$-photon correlated emission by incorporating two distinct
features, (i) surge or rapid emission of photons up to $n$ quanta and (ii) blockade beyond $n$,
which can be applied to any quantum optical systems, not
necessarily anharmonic ones.

\subsection{Photon surge}
Generally, the photo-detection rate $\cal R$ is proportional to the intensity of the optical field under consideration,
${\cal R}\propto\langle{\hat {\cal E}}_-{\hat {\cal E}}_+\rangle$, where the operators ${\hat {\cal E}}_{\pm}$ correspond to the positive- and the negative-frequency part of the field.
Let us assume that a quantum system can be described by a pure steady state $|\Psi\rangle_s$ for simplicity, but our argument applies equally well to mixed states.
If it has emitted $n-1$ quanta, the wavefunction is collapsed to $|\Psi_c^{(n-1)}\rangle={\hat {\cal E}}_+^{n-1}|\Psi\rangle_s$ conditioned on these emissions.
The detection rate for the succeeding $n$th photon is then given by
\begin{eqnarray}
{\cal R}_n\equiv\frac{\langle \Psi_c^{(n-1)}|{\hat {\cal E}}_-{\hat {\cal E}}_+|\Psi_c^{(n-1)}\rangle}{\langle\Psi_c^{(n-1)}|\Psi_c^{(n-1)}\rangle}
=\frac{\langle{\hat {\cal E}}_-^n{\hat {\cal E}}_+^n\rangle}{\langle{\hat {\cal E}}_-^{n-1}{\hat {\cal E}}_+^{n-1}\rangle}
\end{eqnarray}
after the normalization of the conditional state
$|\Psi_c^{(n-1)}\rangle$. Specifically, if the emission out of the
system is a bunch of highly correlated $n$-photons, the second
photon will be emitted right after the first photon and the third
photon after the second, and so on. This idea can be used to
construct our criterion as follows.

The ``bare" rate for the first emission is simply given by the intensity, ${\cal R}_1=\langle{\hat {\cal E}}_-{\hat {\cal E}}_+\rangle$, which only characterizes the signal strength and has little to do with correlation.
For $n$-photon correlation ($n>1$), once a photon is emitted, however, the next emission will immediately follow, thus the conditional rate ${\cal R}_2$ must be large enough. In particular, we require ${\cal R}_2$ to be larger than ${\cal R}_1$, i.e. ${\rm R}_{2,1}\equiv \frac{{\cal R}_2}{{\cal R}_1}=\frac{\langle{\hat {\cal E}}_-^2{\hat {\cal E}}_+^2\rangle}{\langle{\hat {\cal E}}_-{\hat {\cal E}}_+\rangle^2}>1$, which is nothing but the bunching condition in the Glauber $g^{(2)}$ function.
Extending the requirement to next emissions sequentially, we derive a set of {\it surge} conditions
\begin{eqnarray}
{\rm R}_{k,k-1}\equiv\frac{{\cal R}_k}{{\cal R}_{k-1}}=\frac{\langle{\hat {\cal E}}_-^k{\hat {\cal E}}_+^k\rangle\langle{\hat {\cal E}}_-^{k-2}{\hat {\cal E}}_+^{k-2}\rangle}{\langle{\hat {\cal E}}_-^{k-1}{\hat {\cal E}}_+^{k-1}\rangle^2}>1,\hspace{0.5cm}
(k=2,\dots,n),
\label{eqn:ratio}
\end{eqnarray}
which must be satisfied for each $k=2,\dots,n$.

\subsection{Photon blockade}\label{subsect2-2}
However, the satisfaction of Eq.\ (\ref{eqn:ratio}) for all $k=2,\dots,n$ is not sufficient to ensure $n$-photon correlation, and importantly, one must also look at the next occurrences carefully. After the detection of $n$ photons, the succeeding emissions must be suppressed, which can be expressed as
\begin{eqnarray}
{\rm R}_{k,k-1}<1,\hspace{2cm}(k=n+1,\cdots).
\label{eqn:ratio1}
\end{eqnarray}
The fulfillment of all the surge and the blockade conditions in Eqs.\ (\ref{eqn:ratio}) and (\ref{eqn:ratio1}) respectively constitutes our criterion for $n$-photon correlated emission. Note that the condition (\ref{eqn:ratio1}) coincides with the special case of the higher-order antibunching criteria introduced in \cite{Lee1990} for the nonclassicality of photon statistics, rather than the correlation effect .

In our criterion, it is crucial to use the {\it conditional} rates
${\cal R}_k$, rather than the {\it bare} rates $\langle{\hat {\cal
E}}_-^k{\hat {\cal E}}_+^k\rangle$, as the former takes into account the
correlation between adjacent emissions in a stronger sense. However,
the resulting criterion does not require any conditional
measurements. Instead, the quantities ${\rm R}_{k,k-1}$ in Eqs.
(\ref{eqn:ratio}) and (\ref{eqn:ratio1}) simply involve various
photon-coincidence rates and we particularly note that the numerator
and the denominator are in the same order of the field strength. It
is thus given in an experimentally desirable form, that is,
insensitive to the quantum efficiency of photodetectors.

\subsection{Measure of multi-photon correlation}
The conditions in Eqs.\ (\ref{eqn:ratio}) and (\ref{eqn:ratio1}) may be used to define a quantitative measure ${\cal M}_n$ of $n$-photon correlation as
\begin{eqnarray}
{\cal M}_n\equiv \prod_{k=2}^n{\rm max} \{{\rm R}_{kk-1}-1,0\}\prod_{k=n+1}^{N_{\rm tr}}{\rm max} \{{\rm R}_{kk-1}^{-1}-1,0\},
\label{eqn:measure}
\end{eqnarray}
where $N_{\rm tr}$ is a truncated excitation number to be taken
appropriate to a given situation. ${\cal M}_n$ quantifies the
strength of the $n$-photon correlation by measuring the deviation of
${\rm R}_{k,k-1}$ from unity in the surge and the blockade
conditions of Eqs.\ (\ref{eqn:ratio}) and (\ref{eqn:ratio1}),
respectively, and returns a nonzero value only when all those
conditions are satisfied. To experimentally obtain ${\cal M}_n$ for
a given system, one first measures the bare $k$-photon coincidence
rates $\langle{\hat {\cal E}}_-^k{\hat {\cal E}}_+^k\rangle$ for all
$k=1,\dots,N_{\rm tr}$. Then, each conditional rate ${\rm
R}_{k,k-1}$ defined by Eq.\ (\ref{eqn:ratio}) is evaluated and
plugged in to Eq.\ (\ref{eqn:measure}) to determine the value of
${\cal M}_n$.

\subsection{Remarks}
(a) Conventionally, multi-photon correlations have been discussed in terms of the Glauber coherence functions
\begin{eqnarray}g^{(n)}\equiv\frac{\langle{\hat {\cal E}}_-(x_1){\hat {\cal E}}_-(x_2)\cdots{\hat {\cal E}}_-(x_n){\hat {\cal E}}_+(x_n)\cdots{\hat {\cal E}}_+(x_2){\hat {\cal E}}_+(x_1)\rangle}{\langle{\hat {\cal E}}_-(x_1){\hat {\cal E}}_+(x_1)\rangle\langle{\hat {\cal E}}_-(x_2){\hat {\cal E}}_+(x_2)\rangle\cdots\langle{\hat {\cal E}}_-(x_n){\hat {\cal E}}_+(x_n)\rangle}
\nonumber
\end{eqnarray}
($x_i$: a general space-time point) \cite{Glauber}. 
The context of correlation in $g^{(n)}$ is, however, rather limited
and we particularly note that $g^{(n)}$ compares the $n$-photon
coincidence rate (numerator) only with the single-photon counting
rates (denominator). Large (small) value of $g^{(n)}$ characterizes
a bunching (antibunching) effect with no strict $n$-photon
correlation that can emerge even in a classical scattering system,
e.g. $g^{(n)}=n!$ for a thermal light (Hanbury-Brown--Twiss effect
\cite{Hanbury,Assman}). Another example of $g^{(2)}\gg1$ with no
rigorous two-photon correlation will be shown below in
Sec.\ref{sect:cqed}.

(b) It is, therefore, interesting to ask whether our criteria of
$n$-photon correlation can be fulfilled by a classical source. It turns out that, as mentioned in Sec. \ref{subsect2-2}, the blockade condition in Eq.\ (\ref{eqn:ratio1}) is related to the nonclassicality of light fields \cite{Lee1990,
Klyshko1996}. Let us consider the single-mode case in which the field amplitude ${\cal
E}_+$ (${\cal E}_-$) may be replaced by the annihilation
(creation) operator $a$ ($a^\dag$). 
Then, as a special case of Ref. \cite{Lee1990}, one can show that, for a classical source
represented by a positive-definite Glauber-$P$ function,
$P(\alpha)\ge0$, a Cauchy-Schwarz inequality follows as
\begin{eqnarray}
\langle a^{\dag k}a^{k}\rangle\langle a^{\dag k-2}a^{k-2}\rangle
=\int d^2\alpha|\alpha|^{2k}P(\alpha)\int d^2\alpha|\alpha|^{2k-4}P(\alpha)\nonumber\\
\ge\left(\int d^2\alpha|\alpha|^{2k-2}P(\alpha)\right)^2
=\langle a^{\dag k-1}a^{k-1}\rangle^2.
\end{eqnarray}
The violation of the above inequality, which is nothing but the
blockade condition ${\rm R}_{k,k-1}<1$ in Eq.\
(\ref{eqn:ratio1}), is thus a clear signature of nonclassicality.
So our criteria of multiphoton correlation can be fulfiled only by
nonclassical sources. We emphasize that, in the so called multiphoton antibunching criteria
in \cite{Lee1990,
Klyshko1996}, the
focus was made on how to reveal
nonclassicality of the field by a mathematical approach based on
the positive Glauber-$P$ function, thus lacking a clear
interpretation as multiphoton correlation.

\section{Application: Cavity QED system}\label{sect:cqed}

\subsection{Model}To illustrate our criterion, we consider a cavity QED system---one of the well known anharmonic systems that can be implemented in various experimental platforms \cite{Faraon2008,Birnbaum2005,Dayan2008,Fink2008}. A qubit (two-state atom, quantum dot, etc.) is coupled to a single mode field driven by a classical field. For simplicity we investigate the on-resonance case,  $\omega_A=\omega_C\equiv\omega_0$, where $\omega_A$ is the qubit transition frequency and $\omega_C$ the cavity resonance frequency.
The qubit-cavity system at coupling strength $g$ is then described by the Hamiltonian
\begin{eqnarray}
H=\hbar\omega_0 \left( a^{\dagger}a+\frac{1}{2}\sigma_z\right)+i\hbar g(a^{\dagger}\sigma_--a\sigma_+),
\end{eqnarray}
where $\sigma_{\pm}$ and $\sigma_z$ are the Pauli  pseudospin operators. The composite system has the ground state $|0,g\rangle$ with the energy $E_0=0$ and the polaritonic excited states $|\Psi_\pm^n\rangle=\frac{1}{\sqrt{2}}\left(|n,g\rangle\pm |n-1,e\rangle\right)$ with $E_{n,\pm}=n\hbar\omega_0\pm \hbar g\sqrt{n}$ ($n=1,2\dots$) [See Fig.1 (b)].
Therefore, when the system is driven by an external field at frequency $\omega_L$, $n$-photon resonant absorption may occur \cite{Chough2000} at
\begin{equation}
n\hbar\omega_L=E_{n,\pm}=n\hbar\omega_0\pm \hbar g\sqrt{n}.
\end{equation}

In practical situations, the qubit and the cavity field may interact with Markovian environments, which causes dissipation and decoherence to the system.
The global evolution is then governed by the master equation $\dot\rho=\frac{1}{i\hbar}[H_I,\rho]+\gamma (\sigma_-\rho \sigma_+-\frac{1}{2}\sigma_+\sigma_-\rho-\frac{1}{2}\rho\sigma_+\sigma_-)+\kappa (2a\rho a^{\dag}-a^\dag a\rho-\rho a^\dag a)$, where $\gamma$ ($2\kappa$) is the qubit
(cavity) decay rate and the interaction Hamiltonian
$ H_I=\hbar\delta\left(
 a^{\dag}a+\frac{1}{2}\sigma_z\right)+i\hbar g(a^{\dagger}\sigma_--a\sigma_+)+i\hbar{\cal
 E}(a^{\dag}-a),$ with the driving strength ${\cal E}$ and the detuning $\delta\equiv \omega_0-\omega_L$.

\begin{figure}
\centering
\includegraphics[width=4.6in]{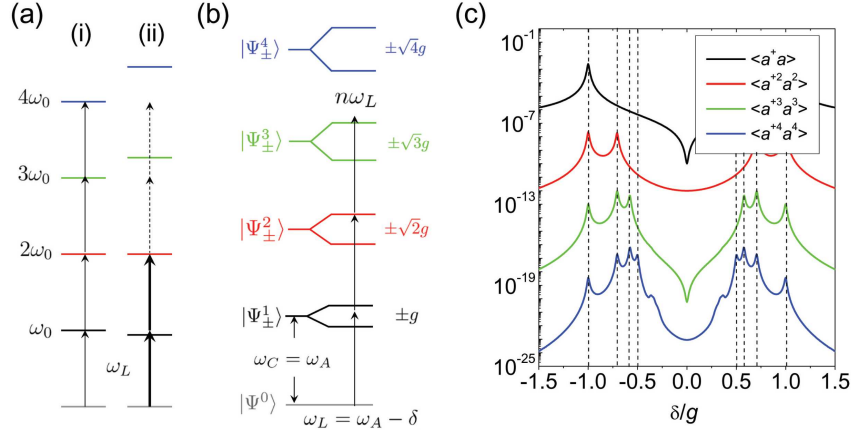}
\caption{(a) Energy-level diagram for (i) harmonic and (ii) anharmonic system. (b) Energy level structure for cavity QED system. (c) multiphoton coincidence rates $\langle a^{\dag n}a^n\rangle$ as a function of $\delta/g$ for  $2\kappa/g=\gamma/g=0.01$ with ${\cal E}/\kappa=0.1$.
The dotted vertical lines represent the locations of the
multiphoton resonances, $\delta=\pm g/\sqrt{n}$ throughout Figs. 1-3.}
\label{fig1}
\end{figure}

\begin{figure}
\centering
\includegraphics[width=4.5in]{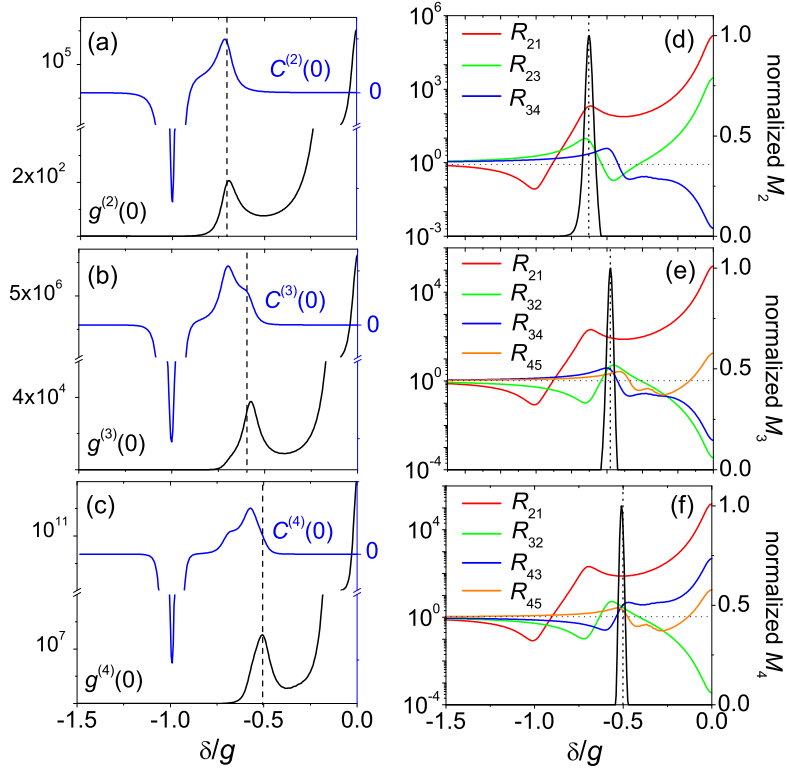}
\caption{(a)-(c) The conventional correlation functions $g^{(n)}(0)$ by Glauber and $C^{(n)}(0)$ by Kubanek {\it et al.} \cite{Kubanek2008}. (d)-(f) The quantitative measure ${\cal M}_n$ together with conditional relative rates ${\rm R}_{k,k-1}$. The truncation numbers used are (d) $N_{\rm tr}=4$, (e) and (f) $N_{\rm tr}=5$. In all plots, $2\kappa/g=\gamma/g={\cal E}/\kappa=0.1$.}
\label{fig2}
\end{figure}

By measuring the cavity transmission as
the driving frequency $\omega_L$ scanned, one may identify the
energy-level structure of the cavity QED system. In Fig.\
\ref{fig1}(c), we plot the bare $n$-photon coincidence rate,
$\langle a^{\dag n}a^n\rangle$ as a function of the normalized
detuning $\delta/g$. In the weak-excitation limit, these rates are
related to the $n$-excitation probability $P_n$ as $\langle a^{\dag
n}a^n\rangle\approx n!P_n$. We see that more resonant peaks are
spectroscopically observed at $\delta=\pm g/\sqrt{n}$ as the order
$n$ is increased in a very strong-coupling regime,
$\gamma/g=2\kappa/g=0.01$. It is important to note that not all the peaks in
$n$th order coincidence rate are relevant to $n$-photon correlation (e.g.,
the peak at $\delta=\pm g$ in the two-photon coincidence), so the
bare coincidence rates may not be used as such to address genuine
multiphoton correlation. To overcome this difficulty, for instance,
one may try to classify those peaks with a prior knowledge on the
excitation paths \cite{Horvath1999}. However, in realistic
situations, the peaks become less resolved as the coupling strength
is reduced (not shown). More importantly, these resonant peaks give
information only on the level structure of excitation and have a
weak connection to correlated emissions.

\subsection{Correlation measures}
Instead, if one measures the Glauber coherence function $g^{(n)}$ of the output, the result may characterize the correlation of emitted photons to some extent, but not in a full rigorous sense.
In particular, $g^{(n)}(0)=\langle
a^{\dagger n}a^n\rangle/\langle a^{\dagger}a\rangle^n$ in Fig.\ \ref{fig2}(a) shows a large bunching at zero detuning $\delta=0$,
which has nothing to do with genuine $n$-photon correlation as we will clearly show below. Close inspection of photon statistics reveals that the system does
exhibit some nonclassical behavior at $\delta=0$, e.g. the
oscillation of conditional detection rate ${\cal R}_k$ which peaks
at even number of $k$, but it is not a rigorous $n$-photon
correlation at any level $n$ in view of our criterion. To get rid of
this ``cumbersome" resonance effect observed at $\delta=0$ that may
overwhelm the other resonance peaks, Kubanek {\it et al.}
introduced the differential correlation function,
$C^{(2)}(0)=\langle a^{\dagger 2}a^2 \rangle-\langle a^{\dagger}a
\rangle^2$, that measures the {\it absolute} occurrence of
two-photon excitation with respect to the single-photon excitation
\cite{Kubanek2008}. The context in this correlation function,
however, is insufficient just like $g^{(2)}(0)$ in general, although it was
instrumental to identify the second resonant peak in
\cite{Kubanek2008}.
Furthermore, a generalization to $n$-photon level, $C^{(n)}(0)=\langle a^{\dagger n}a^n \rangle-\langle a^{\dagger}a\rangle^n$ for $n\ge3$, becomes hardly effective in identifying the higher-order peaks by the broadening effect in the realistic regime [Fig. 2 (a)-(c)].

\begin{figure}
\centering
\includegraphics[width=4.5in]{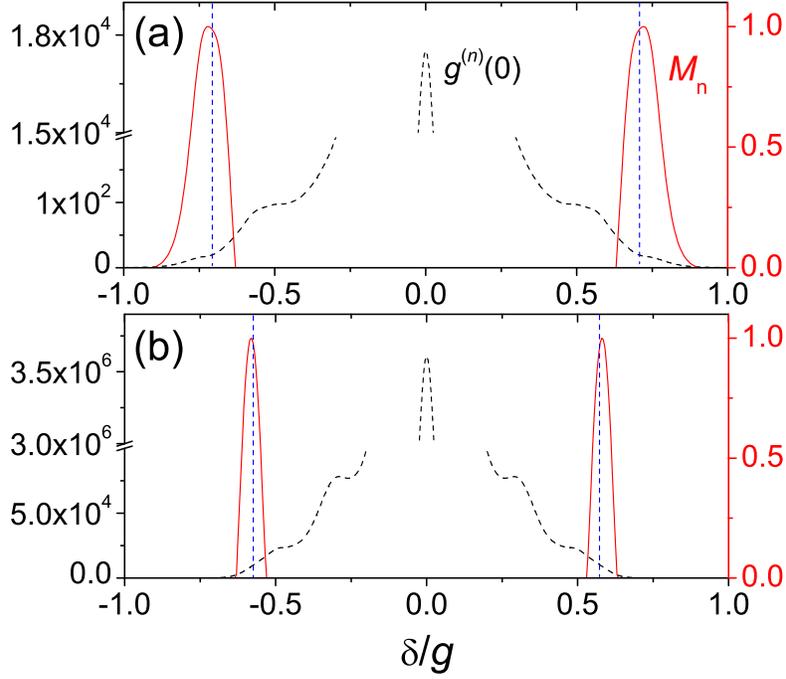}
\caption{Comparison between $g^{(n)}(0)$ (black dotted curve) and our measure ${\cal M}_n$ (red solid curve) for (a) $n=2$ and (b) $n=3$, with the truncation numbers (a) $N_{\rm tr}=4$ and (b) $N_{\rm tr}=5$, respectively. The driving intensity is rather high, ${\cal E}/\kappa=1$, with the coupling condition $2\kappa/g=\gamma/g=0.1$.}
\label{fig3}
\end{figure}

In contrast, our criterion not only detects correlated emission in a well-defined context, but also provides a practical tool to  identify the multi-photon resonance structure of a system in realistic situations.
In Figs.\ \ref{fig2}(d), \ref{fig2}(e), and \ref{fig2}(f), we plot the quantitative measure ${\cal M}_n$ of Eq.\ (\ref{eqn:measure}) for $n=2,3$, and 4 along with various rates ${\rm R}_{k,k-1}$ which are ingredients for constructing the corresponding ${\cal M}_n$. Remarkably, ${\cal M}_n$ yields a positive value only in the spectral vicinity of the resonant peaks $\delta=\pm g/\sqrt{n}$.
The ``spurious" peak at $\delta=0$ disappears by our criterion, which rigorously confirms that this seeming ``resonance" indeed does not represent pure $n$-photon correlated emission.

\subsection{Large driving field}
As we increase the pump strength to obtain more substantial signal, the coincidence spectrum usually becomes difficult to resolve due to the saturation of the system. Our method is, however, still useful for moderately strong pumping owing to the rigorous context established in it. To demonstrate this merit, we have considered the case of a large driving field ${\cal E}/\kappa=1$ in Fig.\ \ref{fig3}, together with the realistic coupling $\gamma/g=2\kappa/g=0.1$. 
Due to the intensity-dependent broadening effect, the Glauber function $g^{(n)}(0)$ no longer shows noticeable marks of resonance, except for the peak at $\delta=0$ overwhelming the entire shape in the spectrum. On the other hand, our measure ${\cal M}_n$ identifies a clear signature of multi-photon correlations under the same condition. This capability would allow one to increase the pump strength to some extent, and thereby easing the difficulty of having to measure higher-order coincidence than $n$ [i.e., blockade conditions in Eq.\ (\ref{eqn:ratio1})] to identify $n$-photon correlation in our method.
Furthermore, we have also checked that other possible broadening effects, e.g. atomic motion in the cavity, do not degrade the capability of our criterion for characterizing multiphoton correlations. We attribute this robustness against experimental imperfections to the rigorous context established with the measure ${\cal M}_n$. 

\section{Conclusion}

In conclusion, we have devised an easy-to-implement criterion for detecting correlated multi-photon emission, imposing surge and blockade requirements in photoemission processes.
A quantitative measure ${\cal M}_n$ has been derived from the correlation context between successive photon emissions in the framework of wavefunction collapse. Our criterion applies to any quantum optical systems, including the ones with anharmonic structure (cavity QED systems, multi-level atoms, etc.).

We have illustrated our method can efficiently detect multi-photon correlations at the resonant peaks of the cavity QED system in contrast to the existing correlation functions.
Note that the anharmonic spectrum which scales as $\sqrt{n}$ is a clear signature of quantum nature of light field \cite{Mossberg,Carmichael1996}, and it thus has been of considerable interest for long but experimentally verified only recently \cite{Schuster2008,Fink2008,Bishop2008}.
In the optical cavity-QED system \cite{Birnbaum2005,Dayan2008,Kubanek2008,Schuster2008}, it becomes harder to directly observe this anharmonicity in higher-orders due to less strong coupling than in the microwave circuit QED system, but our method remarkably makes it possible to clearly pick up the $\sqrt{n}$-dependence despite experimental limitations.
We anticipate that our conceptually rigorous approach can also be useful in addressing correlation effects in other quantum systems beyond optics.

\section*{Acknowledgments}
HN is grateful to H. J. Carmichael for helpful discussions and remarks.
This work was supported by NRL and WCU Grants. HN was supported by the NPRP grant 08-043-1-011 from Qatar National Research Fund.


\begin{thebibliography}{99}

\bibitem{Chang2007}
D.\ E.\ Chang, A.\ S.\ S\o rensen, E.\ A.\ Demler, and M.\
D.\ Lukin, ``A single-photon transistor using nanoscale surface plasmons," Nat. Phys. {\bf 3}, 807-812 (2007)
\bibitem{Faraon2008}
A.\ Faraon, I.\ Fushman, D.\ Englund, N.\ Stoltz, P.\ Petroff, and J.\ Vu\v ckovi\' c, ``Coherent generation of non-classical light on a chip via photon-induced tunnelling and blockade," Nat. Phys. {\bf 4}, 859-863 (2008).

\bibitem{Birnbaum2005}
K.\ M.\ Birnbaum, A.\ Boca, R.\ Miller, A.\ D.\ Boozer, T.\ E.\ Northup, and H.\ J.\ Kimble, ``Photon blockade in an optical cavity with one trapped atom," Nature {\bf 436}, 87-90 (2005).

\bibitem{Dayan2008}
B.\ Dayan, A.\ S.\ Parkins, T.\ Aoki, E.\ P.\ Ostby, K.\ J.\
Vahala, and H.\ J.\ Kimble, ``A photon turnstile dynamically
regulated by one atom," Science {\bf 319} 1062-1065 (2008).

\bibitem{Hartmann2006}
M.\ J.\ Hartmann, F.\ G.\ S.\ L.\ Brand\~ ao, and M.\ B.\ Plenio, ``Strongly interacting polaritons in coupled arrays of cavities," Nat. Phys. {\bf 2}, 849-855 (2006); M.\ J.\ Hartmann and M.\ B.\ Plenio, ``Strong photon nonlinearities and photonic Mott insulators," \prl {\bf 99}, 103601 (2007).

\bibitem{Greentree2006}
A.\ D.\ Greentree, C.\ Tahan, J.\ H.\ Cole, and L.\ C.\ L.\
Hollenberg, ``Quantum phase transitions of light," Nat. Phys. {\bf 2}, 856-861 (2006).

\bibitem{Angelakis2007}
D.\ G.\ Angelakis, M.\ F.\ Santos, and S.\ Bose, ``Photon-blockade-induced Mott transitions and XY spin models in coupled cavity arrays," Phys. Rev. A {\bf
76}, 031805(R) (2007).

\bibitem{Na2008}
N.\ Na, S.\ Utsunomiya, L.\ Tian, and Y.\ Yamamoto, ``Strongly correlated polaritons in a two-dimensional array of photonic crystal microcavities," Phys. Rev. A {\bf
77}, 031803(R) (2008).

\bibitem{Chang}
D.\ E.\ Chang, V.\
Gritsev, G.\ Morigi, V.\ Vuleti$\acute{\textrm{c}}$, M.\ D.\ Lukin,
and E.\ A.\ Demler, ``Crystallization of strongly interacting photons in a nonlinear optical fibre," Nat. Phys.\ {\bf 4}, 884-889 (2008).
%See also T. Kinoshita, T. Wenger, and D. S. Weiss, Science {\bf 305}, 1125 (2004);
%B. Paredes {\it et al.}, Nature {\bf 429}, 277 (2004).

\bibitem{Imamoglu1997}
A.\ Imamoglu, H.\ Schmidt, G.\ Woods,  and M.\ Deutsch, ``Strongly interacting photons in a nonlinear cavity," Phys. Rev. Lett. {\bf 79}, 1467-1470 (1997).

\bibitem{Carmichael1992}
L. Tian and H.\ J.\ Carmichael, ``Quantum trajectory simulations of two-state behavior in an optical cavity containing one atom," Phys. Rev. A {\bf 46}, R6801-R6804 (1992).

\bibitem{Kubanek2008}
A.\ Kubanek, A.\ Ourjoumtsev, I.\ Schuster, M.\ Koch, P.\ W.\ H.\ Pinkse, K.\ Murr, and G.\ Rempe, ``Two-photon gateway in one-atom cavity quantum electrodynamics," Phys. Rev. Lett. {\bf 101}, 203602 (2008).

\bibitem{Schuster2008}
I.\ Schuster, A.\ Kubanek, A.\ Fuhrmanek, T.\ Puppe, P.\ W.\ H.\ Pinkse, K.\ Murr and G.\ Rempe, ``Nonlinear spectroscopy of photons bound to one atom," Nat. Phys.\ {\bf 4}, 382-385 (2008).

\bibitem{Fink2008}
J.\ M.\ Fink, M.\ G\"oppl, M.\ Baur, R.\ Bianchetti, P.\ J.\ Leek, A.\ Blais, and A.\ Wallraff, ``Climbing the Jaynes-Cummings ladder and observing its $\sqrt{n}$ nonlinearity in a cavity QED system," Nature (London) {\bf 454}, 315-318 (2008).

\bibitem{Bishop2008}
L.\ S.\ Bishop, J.\ M.\ Chow, J.\ Koch, A.\ A.\ Houck, M.\ H.\
Devoret, E.\ Thuneberg, S.\ M.\ Girvin, and R.\ J.\ Schoelkopf, ``Nonlinear response of the vacuum Rabi resonance," Nat. Phys.\ {\textbf 5}, 105-109 (2009).

\bibitem{Horvath1999}
L.\ Horvath, B.\ C.\ Sanders, and B.\ F.\ Wielinga, ``Multiphoton coincidence spectroscopy," J. Opt. B: Quantum Semiclassic. Opt. {\bf 1} 446-451 (1999).

\bibitem{Glauber}
R. J. Glauber, ``The quantum theory of optical coherence," Phys. Rev. {\bf 130}, 2529-2539 (1963).

\bibitem{Lee1990}
C. T. Lee, ``Higher-order criteria for nonclassical effects in photon statistics," Phys. Rev. A {\bf 41} 1721-1723 (1990).

\bibitem{Klyshko1996}
D. N. Klyshko, ``Observable signs of nonclassical light
," Phys. Lett. A, {\bf 213} 7-15 (1996).

\bibitem{Carmichael1991}
H.\ J.\ Carmichael, R.\ J.\ Brecha, and P.\ R.\ Rice, ``Quantum interference and collapse of the wavefunction in cavity QED," Opt. Comm. {\bf 82}, 73-79 (1991).


\bibitem{Hanbury}
R. Hanbury-Brown and R. Twiss, ``Correlation between photons in two coherent beams of light," Nature {\bf 177}, 27-29 (1956).

\bibitem{Assman}
%The higher-order bunching effects were very recently demonstrated using a semiconductor microcavity system by
M.\ A\ss mann, F.\ Veit, M.\ Bayer, M.\ van der Poel, and J.\ M.\ Hvam, ``Higher-order photon bunching in a
semiconductor microcavity," Science {\bf 325}, 297-300 (2009).

\bibitem{Chough2000}
Y.-T.\ Chough, H.-J.\ Moon, H.\ Nha, and K.\ An, ``Single-atom laser based on multiphoton resonances at far-off resonance in the Jaynes-Cummings ladder," Phys. Rev. A {\bf 63}, 013804 (2000).

\bibitem{Mossberg}
Y.\ Zhu, D.\ J.\ Gauthier, S.\ E.\ Morin, Q.\ Wu, H.\ J.\ Carmichael, and T.\ W.\ Mossberg, ``Vacuum Rabi splitting as a feature of linear-dispersion theory: Analysis and experimental observations," Phys.\ Rev.\ Lett.\ {\bf 64}, 2499-2502 (1990).

\bibitem{Carmichael1996}
H.\ J.\ Carmichael, P.\ Kochan, and B.\ C.\ Sanders, ``Photon correlation spectroscopy," Phys. Rev. Lett. {\bf 77}, 631-634 (1996).

\end{thebibliography}
\end{document}